\documentstyle[11pt,aaspp4]{article}
\begin{document}

\title{Rapid $UBVRI$ Follow-up of the Highly Collimated Optical
Afterglow of GRB~010222\footnote{Based on observations collected at
the FLWO 1.2-m telescope and the 1.8-m VATT}}

\author{Krzysztof~Z.~Stanek\altaffilmark{2},
Peter~M.~Garnavich\altaffilmark{3}, 
Saurabh~Jha\altaffilmark{2},
Roy~E.~Kilgard\altaffilmark{2}, 
Jonathan~C.~McDowell\altaffilmark{2},
David~Bersier\altaffilmark{2},
Peter~M.~Challis\altaffilmark{2},
Emilio~Falco\altaffilmark{4},
Jason~L.~Quinn\altaffilmark{3}}
 
\author{\tt e-mail: kstanek@cfa.harvard.edu,
pgarnavi@miranda.phys.nd.edu, sjha, rkilgard, jmcdowell, dbersier,
pchallis, falco@cfa.harvard.edu, jquinn@miranda.phys.nd.edu}
 
\altaffiltext{2}{Harvard-Smithsonian Center for Astrophysics, 60
Garden Street, Cambridge MA 02138}
\altaffiltext{3}{Dept. of Physics, University of Notre Dame, 225
Nieuwland Science Hall, Notre Dame IN 46556}
\altaffiltext{4}{Fred L. Whipple Observatory, PO Box 97, Amado, AZ 85645}

\vspace*{-0.1cm}

\begin{abstract}

\vspace*{-0.1cm}

We present the earliest optical observations of the optical
counterpart to the Gamma-Ray Burst (GRB) 010222, obtained with the
Fred L. Whipple Observatory 1.2-m telescope in $UBVRI$ passbands,
starting $3.64\;$hours after the burst (0.4 hours after public
notification of the burst localization).  We also present late
$R$-band observations of the afterglow obtained with the 1.8-m Vatican
Advanced Technology Telescope $\sim 25\;$days after the burst. The
temporal analysis of our data joined with published data indicates a
steepening decay, independent of wavelength, asymptotically
approaching $F_\nu \propto t^{-0.80\pm 0.05}$ at early times ($t\ll
1\;$day) and $F_\nu \propto t^{-1.30\pm 0.05}$ at late times, with a
sharp break at $t_b=0.72\pm 0.10\;$days. This is the second earliest
observed break of any afterglow (after GRB 980519), which clearly
indicates the importance of rapid multi-band follow-up for GRB
afterglow research.

The optical spectral energy distribution, corrected for small Galactic
reddening, can be fit fairly well by a single power-law with $F_\nu
\propto \nu^{-1.07\pm 0.09}$. However, when we fit using our $BVRI$
data only, we obtain a shallower slope of $-0.88\pm 0.10$, in
excellent agreement with the slope derived from our low-resolution
spectrum ($-0.89\pm 0.03$).

The spectral slope and light curve decay slopes we derive are not
consistent with a jet model despite the presence of a temporal
break. Significant host dust extinction with a star-burst reddening
law would flatten the spectral index to match jet predictions and
still be consistent with the observed spectral energy distribution.
We derive an opening angle of $2.1\deg$, smaller than any listed in
the recent compilation of Frail et al. The total beamed energy
corrected for the jet geometry is $4\times 10^{50}$~erg, very close to
the ``standard'' value of $5\times 10^{50}$~erg found by Frail et
al. for a number of other bursts with light-curve breaks.

\end{abstract}


\section{INTRODUCTION}

The BeppoSAX satellite (Boella et al.~1997) has brought a new
dimension to gamma-ray burst (GRB) research, by rapidly providing good
localization of several bursts per year. This has allowed many GRBs to
be followed up at other wavelengths, including X-ray (Costa et
al.~1997), optical (Groot et al.~1997; van Paradijs et al.~1997) and
radio (Frail et al.~1997).  Good positions have also allowed redshifts
to be measured for a number of GRBs (e.g. GRB 970508: Metzger et
al.~1997), providing a definitive proof of their cosmological origin.

The very bright GRB 010222 was detected by BeppoSAX on February
22.30799 UT (Piro~2001) and ranked as second in fluence and third in
flux from all GRBs observed by BeppoSAX (in~'t Zand et al.~2001).

We began the effort to optically monitor the field of GRB 010222,
starting on February 22.4595 UT, i.e. only $3.64\;$hours after the
burst (with $3.2\;$hours of that delay being external to our effort),
using the Fred L. Whipple Observatory (FLWO) 1.2-m telescope.  The
optical counterpart to GRB 010222 was first announced by Henden
(2001a) and we discovered it independently (McDowell et al.~2001). The
optical transient (OT) was easily recognized as a bright
($R\approx18.4$), new object not present in the DSS image at the
position of $\alpha =14^h52^m12^s.55,\;\; \delta = +43^\circ
01{'}06{''}.3\;\;{\rm (J2000.0)}$ (McDowell et al.~2001). The OT
declined by $\sim 0.2\;$mag during about $2\;$hours of the first
night's observations (Henden \& Vrba 2001; Stanek et al.~2001a).
Absorption line systems at $z=1.477, 1.157$ and possibly also at 0.928
were seen in the optical spectrum of GRB 010222 taken within five
hours of the burst with the FLWO 1.5-m telescope (Garnavich et
al.~2001; Jha et al.~2001a), providing a lower limit to the redshift
of the GRB source.  The BeppoSAX NFI follow-up of GRB 010222 started
about $9\;$hours after the burst (Gandolfi~2001) and detected a strong
X-ray afterglow with a position consistent with the optical
transient. The afterglow was also detected in the radio (Berger \&
Frail 2001), in the sub-millimeter (Fich et al. 2001) and in the near
infrared (Di Paola et al. 2001).

Jha et al. (2001b) have presented our spectroscopic data for GRB
010222, and Masetti et al.~(2001b), Lee et al.~(2001) and Sagar et
al.~(2001) have presented results of multi-band optical observations
of the afterglow.  In this paper we discuss rapid photometric $UBVRI$
follow-up of the GRB 010222 afterglow, with particular attention to
the multi-band temporal behavior and broad-band spectral properties of
the GRB OT.

\section{THE PHOTOMETRIC DATA}

Most of our data were obtained with the F.~L.~Whipple observatory
(FLWO) 1.2-m telescope on three nights: 2001 February 21/22, February
22/23 and February 24/25 UT. We used the ``4Shooter'' CCD mosaic
(Szentgyorgyi et al. 2001) with four thinned, back-side illuminated,
AR-coated Loral $2048^2$ CCDs. The camera has a pixel scale of
$0.335\;\arcsec$/pixel and a field of view of roughly $11.5\;\arcmin$
on a side for each chip.  The data were taken in the $2\times2$ CCD
binning mode. We have obtained $N(U,B,V,R,I)=(2,5,2,12,2),N_{tot}=23$
useful images, with exposure times ranging from $60\;$sec to
$1800\;$sec\footnote{$UBVRI$ photometry and our CCD frames are
available through {\tt anonymous ftp} on {\tt cfa-ftp.harvard.edu}, in
the directory {\tt pub/kstanek/GRB010222}, and through the {\tt WWW}
at {\tt http://cfa-www.harvard.edu/cfa/oir/Research/GRB/}.}.

Deep images at late-time were obtained at the 1.8-m Vatican Advanced
Technology Telescope (VATT) on 2001 March 18 and 19 (UT). Twelve
exposures of 900 seconds each were taken in $R$-band over two nights
with an average seeing of 1.2$''$. The VATT CCD was binned $2\times 2$
to provide a scale of 0.4$''$/pixel.

The data were reduced using elements of the photometric data pipeline
of the DIRECT project (Kaluzny et al.~1998; Stanek et al.~1998), based
on the DAOphot PSF-fitting image reduction package (Stetson 1987;
1992). VATT data were shifted and then combined using a minmax
rejection algorithm.

A calibration of the field was obtained by Henden (2001b) in the
$UBVRI$ bands. We obtained an independent $UBVRI$ calibration with the
VATT telescope on 2001 March 19 (UT) using all-sky photometry of
Landolt standard stars (Landolt 1992).  Comparison of our calibration
in the $BVRI$-bands with that of Henden (2001b) and Sagar et
al.~(2001) reveals small differences of about $\pm0.03\;$mag for the
stars in the field. Such small differences do not affect significantly
any of the results in this paper, so for consistency with other
photometric studies of GRB 010222 we adopt the calibration of Henden
(2001b).  For the $U$-band, these differences are larger, with the
VATT-calibrated bright stars in the field being fainter from Henden's
by about $0.06\;$mag. We decided to use VATT calibration for our
$U$-band data.

\section{THE TEMPORAL BEHAVIOR}

\begin{figure}[t]
\plotfiddle{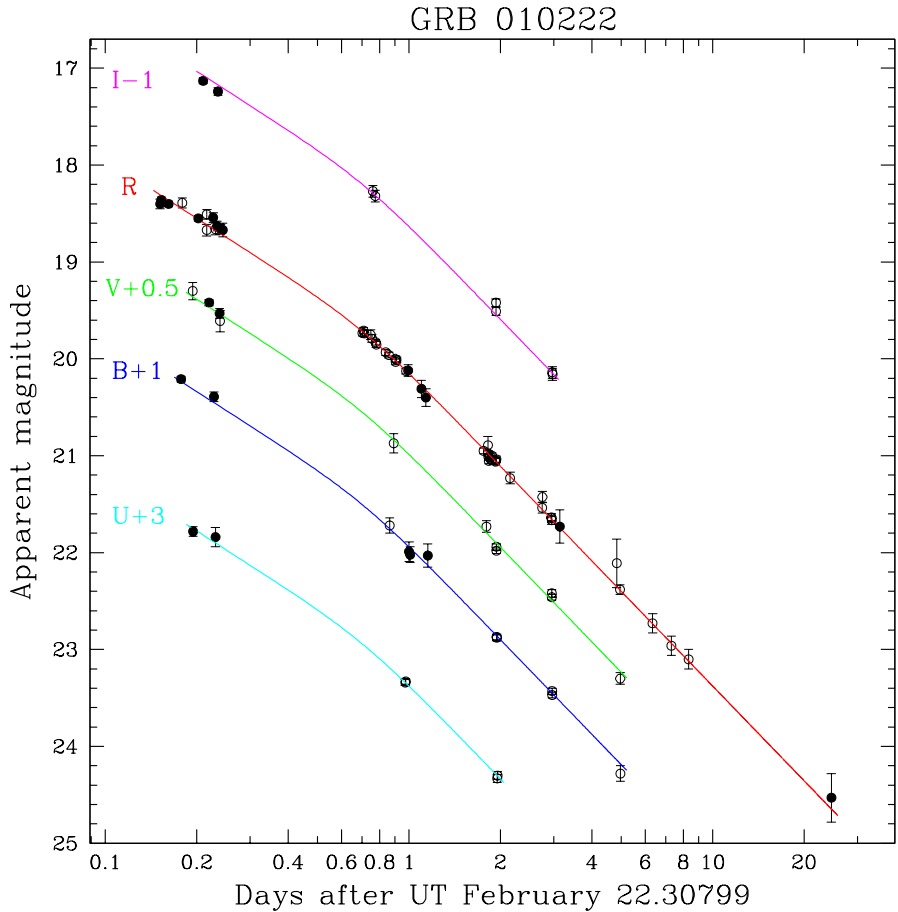}{11cm}{0}{63}{63}{-198}{-105}
\caption{$UBVRI$ light curves of GRB 010222. Our data are shown with
filled circles. Other data shown are those of Billings (2001), Holland
et al.~(2001), Masetti et al.~(2001b) (including their revised
$U$-band) and Veillet(2001a,b). Also shown are the simple analytical
fits discussed in the text.}
\label{fig:time}
\end{figure}

We plot the GRB 010222 $UBVRI$ light curves in Fig.\ref{fig:time}.
Most of the early $UBVRI$ data come from the FLWO 1.2-m telescope
(McDowell et al.~2001; Stanek et al.~2001a), with additional later
data from FLWO (Stanek et al.~2001b; Stanek \& Falco 2001) and from
the VATT 1.8-m telescope (Garnavich et al.~2001b).  To obtain as clear
a picture as possible of the temporal evolution of the afterglow, we
include a few other uniformly reduced data sets into the analysis.
These are $UBVRI$ data of Masetti et al. (2001b) (their $U$-band
magnitudes were revised upwards by about $0.2\;$mag: Masetti, private
communication), $V$-band data of Billings~(2001) reduced by Masetti et
al. (2001b), $R$-band data of Holland et al.~(2001) and $R$-band data
of Veillet~(2001a,b).  To allow for small differences in the reduction
procedures and photometric calibration, uncertainties smaller than
$0.03\;$mag in our and other data were increased to $0.03\;$mag. The
combined data set has the following number of points: $N(U,B,V,R,I)=
(6,11,11,52,8)$, for a total of 88 points.

As noticed by Masetti et al.~(2001a), the optical $R$-band data from
$\sim2$ days after the GRB showed clear departure from the initial
shallow power-law of about $F_\nu \propto t^{-0.9}$ (but see below)
determined by Price et al.~(2001b), Fynbo et al.~(2001) and Stanek et
al.~(2001b).  This trend of a steepening decay was confirmed by Stanek
\& Falco~(2001), Veillet (2001a) and Holland et al.~(2001).

To describe the temporal evolution of the GRB 010222 optical
counterpart, we fit the compiled $UBVRI$ data with the smoothly broken
power-law model of Beuermann et al.~(1999) (see also Sagar et
al.~2000):
\begin{equation}
F_{\nu}(t) =
\frac{2F_{\nu,0}}{\left[\left(\frac{t}{t_b}\right)^{\alpha_1 s}
+\left(\frac{t}{t_b}\right)^{\alpha_2 s}\right]^{1/s}},
\end{equation}
where $t_b$ is the time of the break, $F_{\nu,0}$ is the flux at $t_b$
and $s$ controls the sharpness of the break, with larger $s$ implying
a sharper break.  This formula describes power-law $t^{-\alpha_1}$
decline at early times ($t\ll t_b$) and another power-law
$t^{-\alpha_2}$ decline at late times ($t\gg t_b$). For $s=1$ this
formula converts to the one used by Stanek et al.~(1999) to describe
the behavior of GRB 990510 afterglow. This combined fit has nine free
parameters: five normalization constants $F_{\nu,0}$ (one for each
band) and four shape parameters $\alpha_1,\alpha_2,s,t_b$.  We obtain
the following values for the parameters: $\alpha_1=0.80\pm
0.05,\alpha_2=1.30\pm 0.05, s\approx10(\gg 1),t_b=0.72\pm 0.1\;$day.
These numbers are in good agreement with these derived by Sagar et
al. (2001), but are quite different from those of Masetti et al.
(2001), which can be probably explained by somewhat different data
sets used for fits.  When we ran the combined fit to our data only (24
$UBVRI$ points), we obtained $\alpha_1=0.63\pm 0.15,\alpha_2=1.43\pm
0.15,t_b=0.7\pm 0.1\;$day.

The results of the combined fit are shown as the continuous lines in
the Fig.\ref{fig:time}.  Considering that these data are
inhomogeneous, the combined fit represents very well the overall
temporal behavior of the $UBVRI$ data, with $\chi^2/DOF= 1.11$. It
should be noted that the afterglow has a very blue color, $U-B\approx
-0.6$, compared to most field stars (star ``A'' of McDowell et
al.~2001 has $U-B\approx 0.3$), so special care should be taken when
deriving OT magnitudes in the $U$ filter, which is notorious for
exhibiting strong color terms.

\section{BROAD-BAND SPECTRAL ENERGY DISTRIBUTION}

GRB 010222 is located at Galactic coordinates of
$l=73\arcdeg\!\!.8775, b=60\arcdeg\!\!.8696$. To remove the effects of
the Galactic interstellar extinction we used the reddening map of
Schlegel, Finkbeiner \& Davis (1998, hereafter: SFD).  The SFD
Galactic reddening towards the burst is small, $E(B-V)=0.023$, which
corresponds to expected values of Galactic extinction ranging from
$A_I=0.044$ to $A_U=0.11$, for the Landolt (1992) CTIO filters and the
standard ($R_V = 3.1$) reddening curve of Cardelli, Clayton \& Mathis
(1989).

\begin{figure}[t]
\plotfiddle{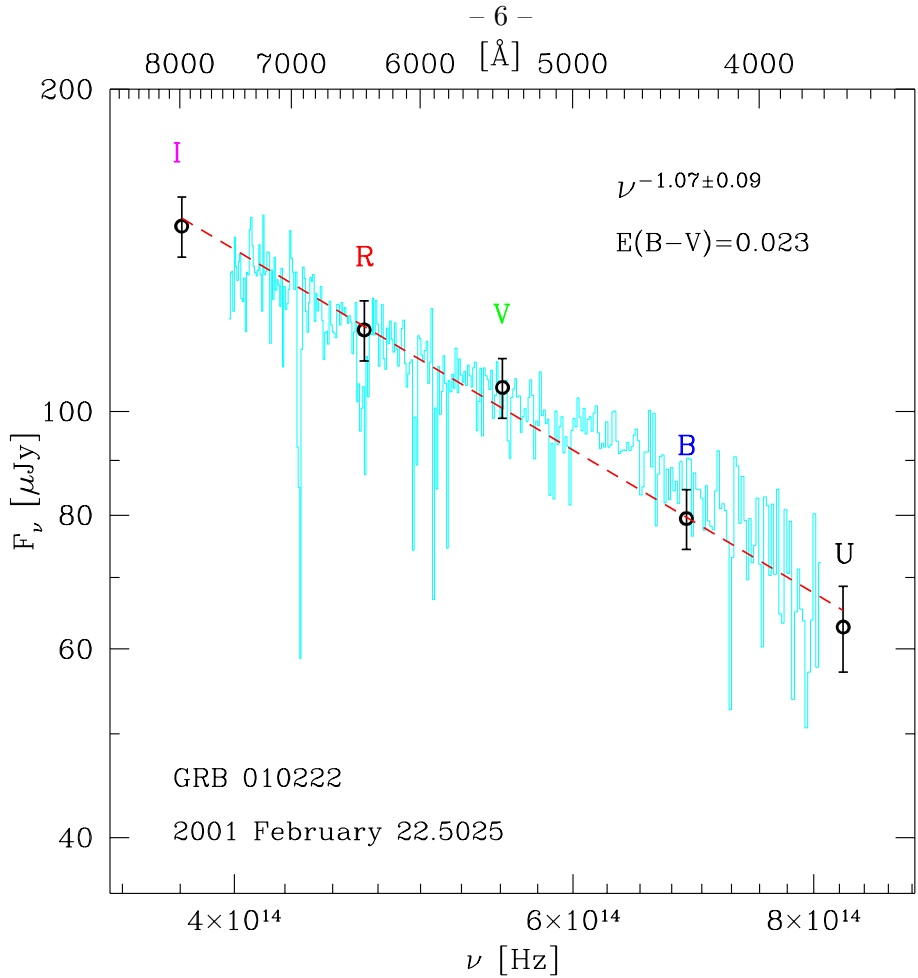}{11cm}{0}{63}{63}{-198}{-105}
\caption{Synthetic spectrum of GRB 010222 $4.7\;hours$ after the
burst, constructed using analytical fits shown in
Fig.\ref{fig:time}.  Also shown is the low-resolution spectrum
taken with the FLWO 1.5-m telescope five hours after the burst (Jha et
al.~2001b), binned to $12\AA$ resolution for
presentation purposes.}
\label{fig:spectrum}
\end{figure}

We synthesize the $UBVRI$ spectrum from our data by interpolating the
magnitudes to a common time.  As discussed in the previous section,
the optical colors of the GRB 010222 counterpart do not show
significant variation.  We therefore select the epoch of February
22.5025 UT ($4.7\;$hours after the burst) for the color analysis,
which coincides with our first $U$-band image and is roughly in the
middle of our first night's data.  We convert the $UBVRI$ magnitudes
to fluxes using the effective frequencies and normalizations of
Fukugita, Shimasaku \& Ichikawa (1995).  These conversions are
accurate to about 5\%, so to account for the calibration and
interpolation errors we assign to each flux a 7\% error (10\% for the
$U$-band). Note that while the error in the $E(B-V)$ reddening value
has not been applied to the error-bars of individual points, we
include it in the error budget of the fitted slope (following SFD we
took the error in $E(B-V)$ to be $0.02\;$mag).

The results are plotted in Fig.\ref{fig:spectrum} for the
dereddened fluxes. The spectrum is fitted reasonably well by a single
power-law with $\nu^{-1.07\pm 0.09}$, somewhat steeper than the
spectral slope derived from our low-resolution spectrum ($-0.89\pm
0.03$: Jha et al. 2001b), although there is a hint of slight downward
curvature towards the blue end of the FLWO 1.5-m spectrum overplotted
in Fig.\ref{fig:spectrum}.  Indeed, when we fit using only our
$BVRI$ data, we obtain a slope of $\nu^{-0.88\pm 0.10}$, in excellent
agreement with Jha et al. (2001b).  Also, the $g'r'i'z'$ spectral
slope of $-0.90\pm 0.03$ found by Lee et al. (2001) agrees very well
with Jha et al. We agree as well with Lee et al. in the slopes derived
when also using $U/u'$ data.

As mentioned above, both in our broad-band spectrum and in the
low-resolution spectrum of Jha et al. (2001b) there is a hint of
downward curvature towards the blue end of the spectrum, which is also
present in Lee et al. (2001) data.  Lee et al. (2001) discuss possible
reasons for such a downturn, including SMC-like extinction in the host
galaxy (possibly close to the GRB).  They find a good fit to their
$u'g'r'i'z'$ broad-band spectrum when fitting a spectrum with an
intrinsic slope of $\beta_0=0.5$ extincted by $A_V=0.19\;$mag of
SMC-like extinction.

\section{DISCUSSION}

We find that GRB~010222 is another example of an optical afterglow
which shows a break in its light curve.  GRB~010222 has the second
earliest observed break (after GRB 980519: Jaunsen et al.~2001) and
one of the most shallow initial decay rates. Here we compare this
burst to other well-observed GRB afterglows.

While the Lorentz factor is large, there is little observational
difference between a spherically expanding shell and a jet.  Only when
the inverse Lorentz factor exceeds the opening angle of the jet does
the difference between an isotropic model and a beamed model become
apparent and this should appear as a break in the afterglow light
curve.  Sari, Piran, \& Halpern (1999) showed that the length of time
between the burst and the afterglow break can be used to estimate the
jet opening angle, although the estimate depends weakly on the burst
fluence and the density of the surrounding gas (which we take to be
0.1~${\rm cm^{-3}}$).  The GRB~010222 fluence in the $40-700$~keV
range of BeppoSAX was $9.2\times 10^{-5}$~${\rm erg\; cm^{-2}}$ (in 't
Zand et al. 2001) at a redshift of $z=1.477$ (Jha et al. 2001b argue
that the $z=1.477$ system is very likely the GRB host galaxy itself),
corresponding to a luminosity distance of $11.5\;$Gpc (assuming
$\Omega_m=0.3,\; \Omega_\Lambda =0.7,\; H_0=65$).  Thus, the isotropic
$\gamma$-ray energy release was $5.9\times 10^{53}$~erg. GRB~010222
had the earliest observed break of any afterglow, occurring at $0.72
\pm 0.1\;$days after the burst, which suggests a highly collimated
jet. The derived opening angle of $2.1\deg$ is indeed smaller than any
listed in the compilation of Frail et al. (2001). The total beamed
energy corrected for the jet geometry becomes $4\times 10^{50}$~erg,
which is extremely close to the ``standard'' value of $5\times
10^{50}\;$erg found by Frail et al. (2001) for a number of other
bursts with light-curve breaks.

\begin{figure}[t]
\plotfiddle{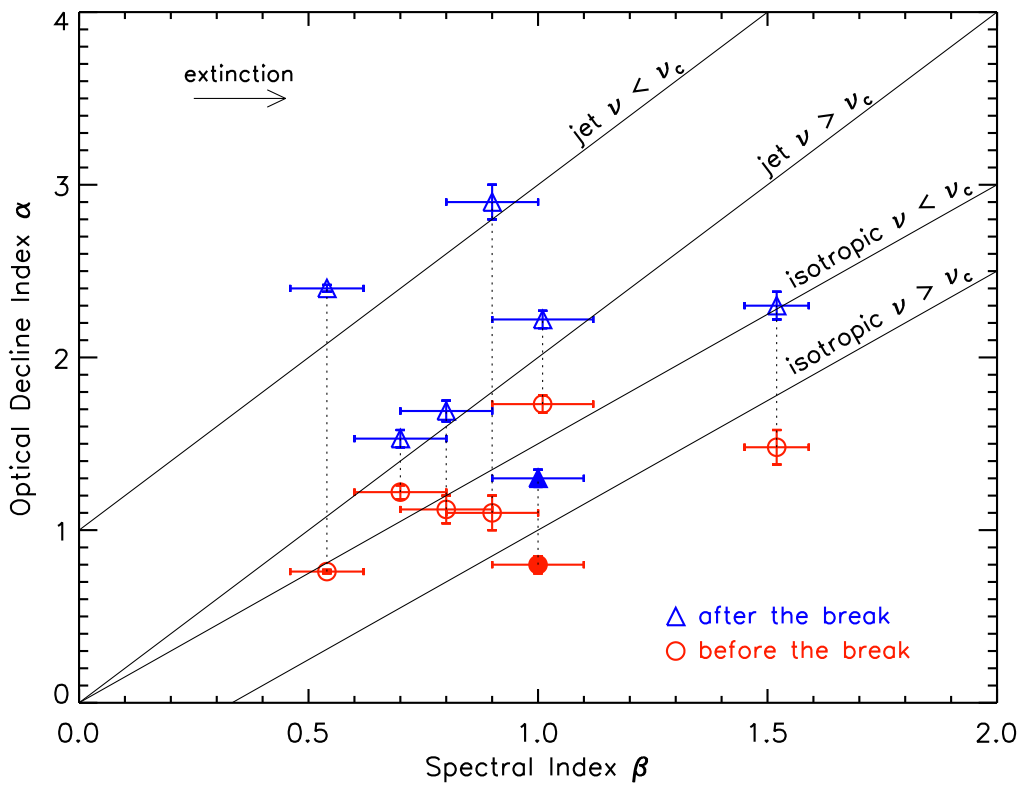}{8.6cm}{0}{70}{70}{-217}{-10}
\caption{A plot of the optical light curve power-law index versus the
spectral power-law index. The solid lines are when the electron
cooling break is at frequencies higher than the optical and the dotted
lines for when the cooling break occurs at frequencies below the
optical. The circles indicate the light curve index $\alpha_1$ before
the temporal break and the stars are the final index $\alpha_2$. Only
well-observed bursts with light curve breaks are shown (in order of
increasing spectral index): GRB~990510 ($z=1.62$), GRB~991216
($z=1.02$), GRB~990123 ($z=1.60$), GRB~000301C ($z=2.03$), GRB~980519,
GRB~000926 ($z=2.04$). GRB~010222 is indicated with solid points.}
\label{fig:beta}
\end{figure}

The optical light curve and spectral indices for the burst are
expected to depend on the electron energy distribution and whether the
observations are made above or below the electron cooling break
frequency. Sari et al. (1999) predict that at frequencies less than
the cooling break, the light curve index, $\alpha$, is related to the
spectral index, $\beta$ by $\alpha_1=3\beta /2$ for an isotropic burst
and $\alpha_2=2\beta +1$ for a jet. Here, $\alpha_1$ denotes the index
before the temporal break, since at that time it is expected to be
indistinguishable from an isotropic burst.  For frequencies above the
cooling break the initial slope is $\alpha_1= 3\beta /2-1/2$ and the
final slope is simply $\alpha_2= 2\beta$.  We can plot these relations
in an $\alpha$ versus $\beta$ space as four lines with the slope for a
beamed burst being steeper than that of an isotropic afterglow.  We
expect the index before the temporal break to fall on one of the two
isotropic models and indeed, when we place a number of well-observed
afterglows with temporal breaks on the plot (Fig.\ref{fig:beta}),
$\alpha_1$ tends to fall near the isotropic models and $\alpha_2$
along the jet models.  Well-observed bursts with light curve breaks
are shown: GRB~990123 (Holland et al.~2000), GRB~990510 (Stanek et
al.~1999), GRB~991216 (Garnavich et al. 2000a; Halpern et al. 2000),
GRB~000301C (Garnavich et al. 2000b; Rhoads \& Fruchter 2001),
GRB~980519 (Jaunsen et al.~2001), GRB~000926 (Price et al. 2001a). We
do not include GRB~991208 (Castro-Tirado et al. 2001), which is well
observed until the `break' defined by only one point.

GRB~000926 (the rightmost pair of points in the figure) does not fit
very well on the plot as the light curve index after the break is much
too small for the Sari et al. (1999) jet model. Price et al. (2001a)
suspect that extinction could explain the mismatch between the
temporal and spectral slopes of this object. Extinction will steepen
the spectral slope, moving the points to the right in
Fig.\ref{fig:beta}, while leaving the light curve parameters
unaffected as long as there is no significant color change in the
afterglow.

For GRB~010222, the final temporal slope is also too small for jet
models given the observed $\beta$. The model can be saved if the
spectral slope was intrinsically in the range $0.5<\beta <0.7$ and has
been steepened by host extinction. Indeed, Lee et al. (2001) find a
good fit to their $u'g'r'i'z'$ broad-band spectrum when fitting an
intrinsic slope of $\beta_0=0.5$ extincted by $A_V=0.19\;$mag of
SMC-like extinction (their Fig.2).

\begin{figure}[t]
\plotfiddle{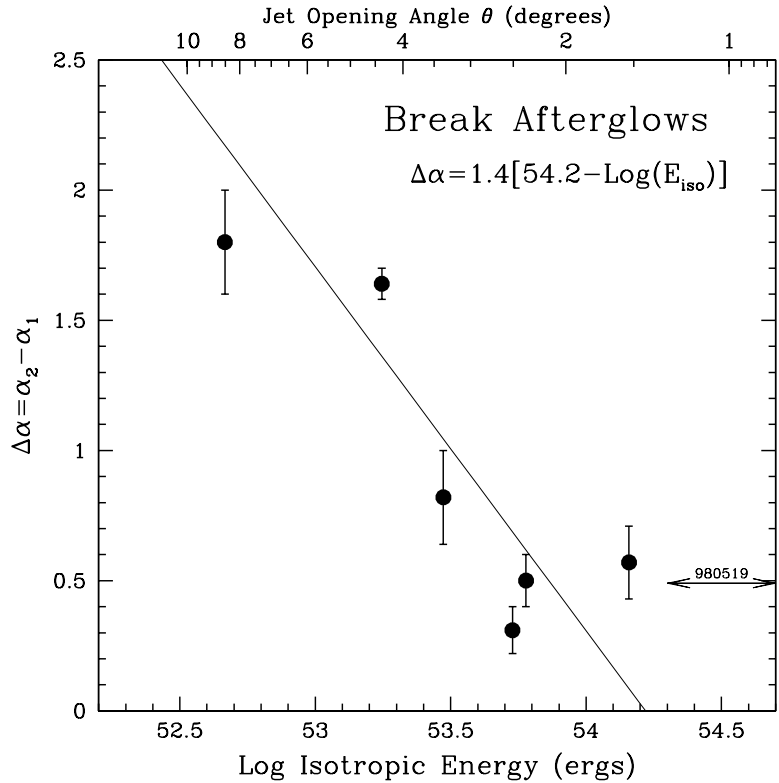}{8.6cm}{0}{55}{55}{-178}{-98}
\caption{The amplitude of the light curve break
$\Delta\alpha=\alpha_2-\alpha_1$ versus the equivalent isotropic
energy release of the burst.  GRB~980519, for which there is no
redshift measured, is indicated with an arrow. The jet opening angle
on the upper axis is given by $\theta=\sqrt{2 E_0/E_{iso}}$, where
$E_0=10^{50.7}\;$erg is the ``standard'' energy of the GRB found by
Frail et al.~(2001).  The line shows a weighted linear fit to the
points which has a slope of 1.4$\pm 0.1$.  The correlation, however
uncertain, seems to indicate that bursts which are more highly
collimated show a smaller amplitude break.}
\label{fig:eiso}
\end{figure}

The magnitude of the shift in light curve slope across the break,
$\Delta\alpha=\alpha_2-\alpha_1$, is independent of extinction to
first order so it can provide clues to the break process without the
uncertainty introduced by dust.  The size of the shift depends on the
electron energy distribution, but Panaitescu \& Kumar (2000) point out
that the spectral cooling break moving through the optical band could
also cause or contribute to a light curve break.  The density
distribution of the circumburst environment further influences the
break amplitude making it difficult to sort out all the possible
combinations of parameters.  We plot the change in slope versus the
equivalent isotropic energy release of the burst in
Fig.\ref{fig:eiso}.  For an interstellar medium an electron cooling
break alone could only change the slope by $\Delta\alpha < 0.25$ which
is inconsistent with all but one burst. But, for a circumstellar gas
distribution the cooling break can change the slope by as much as 1.25
for large values of the spectral index. This slope change is still too
small to explain two observed bursts requiring that at least some of
the events be collimated.

The number of well-observed break afterglows is still small, making
any conclusions about Fig.\ref{fig:eiso} somewhat foolhardy.  A
majority of the break bursts have $\Delta\alpha\sim 0.5\pm0.2$ while
two of the bursts have significantly larger break amplitudes of
$\Delta\alpha > 1.5$. There may be a mild correlation with equivalent
isotropic energy which suggests that the smallest jump in light curve
index occurs at the highest isotropic energies.  Why this apparent
correlation may exist is unclear, but given the relation between
isotropic energy and jet opening angle suggested by Frail et
al. (2001), the break amplitude may be related to the geometry of the
burst.

\section{CONCLUSIONS}

We presented rapid $UBVRI$ observations of GRB 010222, starting
$3.64\;$hours after the burst. Analysis of the data indicates a
steepening decay, independent of optical wavelength, with power-law
behavior $F_\nu \propto t^{-0.80\pm 0.05}$ at early times ($t\ll
1\;$day) and second power-law $F_\nu \propto t^{-1.30\pm 0.05}$ at
late times, with the break time at $t_0=0.72\pm 0.1\;$days. This is
yet another example of such a broad-band break for a GRB OT and it is
very well documented. The broad-band spectral slope and early break
time stress the importance of multi-band optical observations for GRB
studies, especially early after the burst.

The spectral slope and light curve decay slopes we derive are not
consistent with the Sari et al. (1999) jet model despite the presence
of a temporal break. As shown by Lee et al. (2001), significant host
dust extinction with a star-burst reddening law would flatten the
spectral index by as much as 0.5 and match predictions of a jet
geometry.  Such extinction would still be consistent with our observed
spectral energy distribution.  We derive an opening angle of
$2.1\deg$, smaller than any listed in the recent compilation of Frail
et al. (2001). The total beamed energy corrected for the jet geometry
is very close to the `standard' energy value found by Frail et al. for
a number of other bursts with light-curve breaks.

We find a correlation between the estimated isotropic energy release
and the change in power-law index across the light curve break with
the higher energy bursts showing the smallest change in light curve
slope.  It is not clear why this correlation might exist, but the
relation between isotropic energy and jet opening angle suggests that
the jet geometry may be involved, with bursts which are more highly
collimated showing a shallower break.

\acknowledgments{We thank the BeppoSAX team, Scott Barthelmy and the
GRB Coordinates Network (GCN) for the quick turnaround in providing
precise GRB positions to the astronomical community.  We thank Nicola
Masetti for exchanging data with us. We also thank Tom Matheson and
Mike Pahre for valuable discussions and the anonymous referee for
useful comments. PMG acknowledges support from the NASA LTSA grant
NAG5-9364.}

\end{document}